# Bi-chromatic paraxial beam as a model of spatio-temporal light fields

A. Y. Bekshaev

[1]*Physics Research Institute, I.I. Mechnikov National University, Dvorianska 2, 65082, Odessa, Ukraine*
*bekshaev@onu.edu.ua*

**Abstract:** Optical fields with rich and well-developed spatio-temporal structure, including ultra-short structured light pulses, are essentially non-monochromatic and contain a continuous spectrum of monochromatic constituents. However, some substantial features of such fields and physical mechanisms determining their behavior can be understood based on simplified models including only two monochromatic paraxial components. We consider examples of such model beams, their specific spatial and temporal properties as well as their descriptive abilities for the meaningful characterization of realistic spatio-temporal light fields. In particular, the proposed model enables an explicit consistent analysis of the photon-probability distributions in non-monochromatic fields, which confirms a high degree of coincidence between the "energy center" and "probability center" of the field. Simultaneously, particular features of the two-component bi-chromatic paraxial fields (periodic and rotational character of the longitudinal and temporal evolution, specific deformations of the propagating-beam transverse intensity profile, etc.) are inspected using numerical examples.

## 1. Introduction

In the past decades, a lot of research efforts has been made to the studies of ultrashort light pulses characterized by the complex spatial and temporal structure [1–10]. The specific features of such light fields, e.g., their vortex properties in the spatial of spatio-temporal domain [4–11], not only constitute an academic interest but open new attractive prospects in the optical micro- and nanotechnologies [12–17].

A characteristic feature of the ultrashort pulses is their limited duration in time, from several to several thousand of light-oscillation periods. This fact makes impossible their description based on the monochromatic concepts usual in paraxial light optics, and requires explicit consideration of the continuous temporal spectra of such light fields, the range of optical frequencies $\delta\omega$ being correlated with the pulse duration $\delta t \sim 2\pi/\delta\omega$. Generally speaking, this is not a hard task [1–11] but it is desirable to construct a simplified model of the light pulse propagation and transformation employing the usual picture of monochromatic paraxial beam evolution [18–20]. In particular, such simplified models can be useful for heuristic clarification of the transformations occurring to the 2D transverse profiles of the propagating pulse beams, for the development of pictorial ideas on their succinct characterization.

For example, interesting questions appear relating the descriptive abilities and comparative meanings of the light energy distribution versus the "photon density" distribution [9,10,21]. Their difference appears because paraxial beams of the same amplitude but different frequencies contain different numbers of photons [21]; additionally, the consistent analysis of the photon density in 3D space is hampered by the fact that the photon wave function and the corresponding probability density are generally ill-defined in real space [21]. Moreover, the distributions of the energy density and photon probability differ in the general relativistic case [22,23].

In this situation, the concept of a polychromatic paraxial beam containing a limited number of discrete frequency components [24–27], may be attractive. In the simplest case, the paraxial model of a spatio-temporal light pulse may include only two discrete components with frequencies $\omega_1$ and $\omega_2$ such that $(\omega_1 + \omega_2)/2$ coincides with the pulse central frequency, and $|\omega_1 - \omega_2| \sim 2\pi/\delta t$. Of course, this model cannot reflect the temporal features of the pulse evolution but it can be useful for understanding the "transverse" spatial field behavior if one considers the properly chosen time interval of the size $\sim 2\pi/|\omega_1 - \omega_2|$ or, equivalently, the beam longitudinal segment corresponding to the "pulse length" $\sim 2\pi/|k_1 - k_2|$ where $k_{1,2} = \omega_{1,2}/c$, $c$ is the light velocity. Besides, based on this model, we intend to consistently analyze the specific relations between the "energy density" and the "probability density" in paraxial polychromatic beams, in particular, the integral pulse-position characteristics: "energy center" and "probability center" [9,10].

The main results of the paper are obtained using the simple superposition of two lowest Laguerre-Gaussian (LG) modes of different frequencies (Sections 2, 3); from now on, it will be referred as "model beam" (MB). Based on the MB properties, the influence of the spectrum width is analyzed by varying the frequency difference of the components. Simultaneously, we recognize that our MB is analogous to the known family of "rotating beams", formed by oppositely charged vortex LG modes [18,19] with different frequencies, and extensively studied previously [24–27]. During the detailed inspection of the MB properties (Section 4), we cannot omit the comparison of MB with known rotating beams, which enables to disclose their similarities as well as important distinctive features.

## 2. General expressions

Let us consider a linearly polarized paraxial wave packet (longitudinal axis of propagation coincides with the axis $z$) where the transverse electric field is directed along the axis $x$, and the transverse magnetic field is then $y$-directed. In general, the instantaneous electric field can be presented in the form [1,2,11]

$$E_x = \text{Re} \int u(x, y, z, \omega) \exp[i(kz - \omega t)] \frac{d\omega}{2\pi} \tag{1}$$

where $k$ is the monochromatic component wavenumber, $\omega = ck$ means its frequency; the spectral amplitude $u(x, y, z, \omega)$ is supposed to practically vanish outside the spectral range of the width $\delta\omega$ with the central frequency $\omega_0$. Here and in the rest of the paper, absence of the integral limits implies the infinite limits, and the integrand behavior is supposed to warrant the integral convergence.

To model the general situation, we reduce the continuous spectrum of (1) to the discrete set of two frequencies $\omega_{1,2}$; then, the field under consideration is described by a superposition of two paraxial beams, $E_1$ and $E_2$ with the wavenumbers $k_{1,2}$. Accordingly, instead of (1), the electric and magnetic fields of the superposition acquire the form [18–20]:

$$E_x = \text{Re}(E_1 + E_2), \quad H_y = \text{Re}(H_1 + H_2) \tag{2}$$

where the complex-valued fields $E_{1,2}$ and $H_{1,2}$ can be expressed through the slowly-varying complex amplitudes $u_{1,2}$:

$$E_{1,2}(x, y, z, t) = H_{1,2}(x, y, z, t) = u_{1,2}(x, y, z) \exp(ik_{1,2}s), \tag{3}$$

$s = z - ct$. Equations (2) and (3) describe the model superposition which is the subject of further investigation.

The first task is to represent the spatial distributions of the field energy and related quantities in the considered two-frequency superposition. As is known [25], in paraxial beams the field energy density $\tilde{w}_e$ and the longitudinal energy flow density $\tilde{S}_z$ can be expressed as

$$\tilde{S}_z = c\tilde{w}_e = \frac{c}{8\pi}\left(E_x^2 + H_y^2\right) \tag{4}$$

(the Gaussian system of units is employed). In view of Eqs. (3), these quantities contain time-independent terms, slow-varying terms oscillating with frequency $|\Delta\omega| = |\omega_2 - \omega_1|$, and the fast-varying terms (oscillation frequencies $2\omega_1$, $2\omega_2$ and $\omega_1 + \omega_2$). Under usual conditions, the fast oscillations are unobservable, and their contributions in Eq. (4) should be time-averaged, whereas the slow terms are kept. After this operation [25], the time-averaged expressions for the energy $w_e$ and energy flow $S_z$ densities can be presented in the form

$$w_e = \frac{1}{c} S_z = \frac{1}{8\pi} |E_1 + E_2|^2 = \frac{1}{8\pi} \left( |u_1|^2 + |u_2|^2 + u_1^* u_2 e^{i\Delta k \cdot s} + u_2^* u_1 e^{-i\Delta k \cdot s} \right). \tag{5}$$

where $\Delta k = k_2 - k_1 = \Delta\omega / c$. These distributions describe the optical-field intensity patterns in any point of space. Additionally, the distributions (5) can be succinctly characterized by some parameters, the main of which is the energy center (EC) determined as [19]:

$$\begin{pmatrix} x_e \\ y_e \end{pmatrix} = \frac{1}{W_e} \int \begin{pmatrix} x \\ y \end{pmatrix} w_e \, dxdy = \frac{1}{8\pi W_e} \int \begin{pmatrix} x \\ y \end{pmatrix} |E_1 + E_2|^2 \, dxdy \tag{6}$$

where

$$W_e = \int w_e \, dxdy = \frac{1}{8\pi} \int |E_1 + E_2|^2 \, dxdy \tag{7}$$

is the linear energy density per unit length $z$.

Another important distribution is the "photon density", or "probability distribution" [9,10]. For a monochromatic field, the probability density $w_p$ is proportional to the energy density $w_e$ divided by the photon energy $\hbar\omega = \hbar c k$, that is, $w_p = |\psi_p|^2 \propto w_e / \omega \propto w_e / k$, where the "photon wave function" is $\psi_p \propto E / \sqrt{k}$ [9,10]. In case of the bi-chromatic superposition (2), (3), considered in this paper, $\psi_p \propto E_1 / \sqrt{k_1} + E_2 / \sqrt{k_2}$; for convenience of comparison with $w_e$, we normalize the probability density $w_p = |\psi_p|^2$ in such a way that, for the monochromatic case $k_1 = k_2$, it reduces to the energy density. As a result, it obtains the form

$$w_p = \frac{1}{8\pi} \sqrt{k_1 k_2} \left| E_1 / \sqrt{k_1} + E_2 / \sqrt{k_2} \right|^2$$

$$= \frac{1}{8\pi} \left( |u_1|^2 \sqrt{\frac{k_2}{k_1}} + |u_2|^2 \sqrt{\frac{k_1}{k_2}} + u_1^* u_2 e^{i\Delta k \cdot s} + u_2^* u_1 e^{-i\Delta k \cdot s} \right). \tag{8}$$

Then, similar to the EC (6), the "probability center" (PC) is determined by equation

$$\begin{pmatrix} x_p \\ y_p \end{pmatrix} = \frac{1}{W_p} \int \begin{pmatrix} x \\ y \end{pmatrix} w_p \, dxdy = \frac{\sqrt{k_1 k_2}}{8\pi W_p} \int \begin{pmatrix} x \\ y \end{pmatrix} \left| \frac{E_1}{\sqrt{k_1}} + \frac{E_2}{\sqrt{k_2}} \right|^2 dxdy \tag{9}$$

where

$$W_p = \int w_p \, dxdy = \frac{\sqrt{k_1 k_2}}{8\pi} \int \left| \frac{E_1}{\sqrt{k_1}} + \frac{E_2}{\sqrt{k_2}} \right|^2 dxdy. \tag{10}$$

Note that Eqs. (6) and (9) determine, normally, $z$-dependent EC and PC characterizing the field in a fixed cross section. For general spatio-temporal pulses with finite longitudinal length, the corresponding analogs of Eqs. (6) and (9) should relate to the whole field, including the longitudinal coordinate $z$ [8,23] (with associated difficulties of the probability density formulation in the coordinate space [10,21]).

## 3. Description of the model beam parameters

For the detailed studies, we suppose that both beams of the superposition (2), (3) are the lowest-order circular LG modes [18,19] whose complex amplitude distributions are described by equations

$$u_1(x,y,z) = \frac{A_1}{b_1} \exp\left(-\frac{x^2+y^2}{2b_1^2} + ik_1\frac{x^2+y^2}{2R_1} - i\arctan\frac{z}{z_{R1}}\right), \quad (11)$$

$$u_2(x,y,z) = \frac{A_2}{b_2}\frac{x+iy}{b_2} \exp\left(-\frac{x^2+y^2}{2b_2^2} + ik_2\frac{x^2+y^2}{2R_2} - 2i\arctan\frac{z}{z_{R2}}\right); \quad (12)$$

the longitudinal coordinate $z$ is counted from the waist plane which is supposed common for both components (11) and (12), and the beam waist radius of both beams is $b_0$, so that the parameters of the distributions (11) and (12) evolve with $z$ according to relations

$$b_{1,2} = b_0\sqrt{1+\frac{z^2}{z_{R1,2}^2}}, \quad R_{1,2} = \frac{z^2 + z_{R1,2}^2}{z}, \quad (13)$$

where

$$z_{R1,2} = k_{1,2}b_0^2. \quad (14)$$

The expressions (11) – (14), together with the Eqs. (2) – (10) above, provide an exhaustive description of the MB considered in the rest of the paper, and are used for its detailed characterization.

First, let us consider the specific features of the integral beam characteristics (6), (7) and (9), (10). Due to the symmetry of expressions (11) and (12), the terms describing interference between the components give vanishing contributions to the integrals (7), (10), and those can be rewritten in the forms

$$W_e = \frac{1}{8\pi}\int\left(|u_1|^2 + |u_2|^2\right)dxdy = \frac{1}{8}\left(|A_1|^2 + |A_2|^2\right),$$

$$W_p = \frac{1}{8\pi}\int\left(|u_1|^2\sqrt{\frac{k_2}{k_1}} + |u_2|^2\sqrt{\frac{k_1}{k_2}}\right)dxdy = \frac{1}{8}\left(|A_1|^2\sqrt{\frac{k_2}{k_1}} + |A_2|^2\sqrt{\frac{k_1}{k_2}}\right). \quad (15)$$

In turn, in the integrals for the EC (6) and PC (9), contributions of the separate beam components, proportional to $|u_1|^2$ and $|u_2|^2$, vanish whereas the interference terms are crucial, which results in

$$\begin{pmatrix}x_{e,p}\\y_{e,p}\end{pmatrix} = \frac{1}{W_{e,p}}\int\begin{pmatrix}x\\y\end{pmatrix}\left(u_1^*u_2 e^{i\Delta k \cdot s} + u_2^*u_1 e^{-i\Delta k \cdot s}\right)dxdy. \quad (16)$$

As is seen from (5) – (10) and (16), the EC and PC positions are functions of the longitudinal coordinate $z$ and $s = z - ct$. Generally, this means that $x_{e,p}(z,t)$ and $y_{e,p}(z,t)$ may exhibit concerted oscillations with the temporal and spatial periods dictated by the "slow" temporal and spatial frequencies $\Delta\omega$ and $\Delta k$, modulated by the gradual change of the beams' parameters described via Eqs. (13), (14). Note that according to (16), the spatio-temporal variations of the MB EC and PC are, generally, identical, and only their magnitudes may differ due to different denominators $W_e$ and $W_p$ (15), so that

$$\frac{x_e}{x_p} = \frac{y_e}{y_p} = \frac{W_p}{W_e} = \frac{\left(|A_1|^2\sqrt{\frac{k_2}{k_1}} + |A_2|^2\sqrt{\frac{k_1}{k_2}}\right)}{|A_1|^2 + |A_2|^2}. \quad (17)$$

Also, we will need the transverse momentum density of the MB field determined as the transverse component of the time-averaged instant Poynting vector $\tilde{\mathbf{P}} = \frac{1}{4\pi c}\mathbf{E}\times\mathbf{H}$ [25]. It originates from the longitudinal components which are inherent in any paraxial field [19], and for the beam of Eqs. (2), (3) equal to

$$E_z = \mathrm{Re}(E_{1z} + E_{2z}), \quad H_z = \mathrm{Re}(H_{1z} + H_{2z})$$

where [19,25]

$$E_{1z,2z} = \frac{i}{k_{1,2}}\frac{\partial u_{1,2}}{\partial x}\exp(ik_{1,2}s), \quad H_{1z,2z} = \frac{i}{k_{1,2}}\frac{\partial u_{1,2}}{\partial y}\exp(ik_{1,2}s).$$

Accordingly, the transverse part of the instant Poynting vector can be found as

$$\tilde{\mathbf{P}}_\perp = \frac{1}{4\pi c}\left[-\mathbf{e}_x(E_{1z}+E_{2z})(H_{1y}+H_{2y}) - \mathbf{e}_y(E_{1x}+E_{2x})(H_{1z}+H_{2z})\right]$$

whence, after omitting the "rapidly oscillating" terms proportional to $\exp(\pm 2ik_{1,2}s)$ and $\exp[\pm i(k_1+k_2)s]$, one obtains the time-average expression

$$\mathbf{P}_\perp = \frac{1}{8\pi c}\mathrm{Re}\left(\frac{i}{k_1}u_1\nabla_\perp u_1^* + \frac{i}{k_2}u_2\nabla_\perp u_2^* + \frac{i}{k_2}u_1\nabla_\perp u_2^* e^{-i\Delta k \cdot s} + \frac{i}{k_1}u_2\nabla_\perp u_1^* e^{i\Delta k \cdot s}\right) \quad (18)$$

($\nabla_\perp = \mathbf{e}_x(\partial/\partial x) + \mathbf{e}_y(\partial/\partial y)$, $\mathbf{e}_x$, $\mathbf{e}_y$ being the unit vectors of the transverse coordinates). In turn, Eq. (18) determines the longitudinal orbital angular momentum (OAM) [18,19] of the MB as

$$L_z = \int(xP_y - yP_x)dxdy. \quad (19)$$

## 4. Numerical analysis

### 4.1. General conventions

Main conclusions on the MB properties and their dependence on the superposition parameters will be derived from the numerical analysis of the above equations. The calculations are performed for the following values of the MB parameters in Eqs. (11) – (14):

$$k_1 = 10^5 \text{ cm}^{-1}, \quad b_0 = 0.01 \text{ cm}, \quad A_1 = A_2 = A = 1 \text{ statV} \quad (\mathrm{Im}A = 0). \quad (20)$$

The accepted value of $k_1$ approximately corresponds to the He-Ne laser radiation, the Gaussian envelope waist radius $b_0$ of (20) determines the Rayleigh length (14) $z_{R1} = 10$ cm; the linear density of the field energy (7) equals to

$$W_e = \frac{1}{4}A^2. \quad (21)$$

The influence of the frequency difference between the components $E_1$ and $E_2$ (2), (3) will be inspected by considering variable values of $k_2$ ($\omega_2$).

To get a reference point, we first reproduce the main features of the monochromatic case when $k_2 = k_1$ in Eqs. (11) – (14) [28]. In this case, the beam evolution can be easily described analytically, and the numerical illustrations are given in Figs. 1, 2. Upon conditions (20), the beam intensity pattern with asymmetric bright spot is formed. In the waist plane, the bright peak is situated at the positive branch of axis $x$ but with growing propagation distance $z$ it moves in the $y$ direction. Simultaneously, the whole transverse distribution spreads due to the beam divergence, and the whole picture looks as a rotation around the $z$-axis (Fig. 1); in the far field, the beam profile is turned by 90°, and the bright peak tends to the $y$-axis [18,28]. This sort of the intensity pattern evolution directly follows from Eq. (5) with account for (11) – (13) with $\Delta k = 0$.

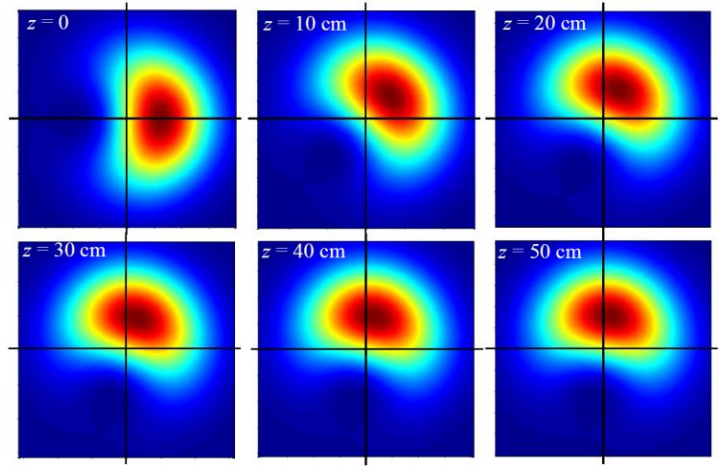

Fig. 1. Transverse intensity distributions of the MB (11), (12) at different distances from the waist plane calculated upon the conditions (20) for the monochromatic case $k_2 = k_1$. The size of each image is $2b_1(z) \times 2b_1(z)$, $b_1(z)$ is determined by Eq. (13).

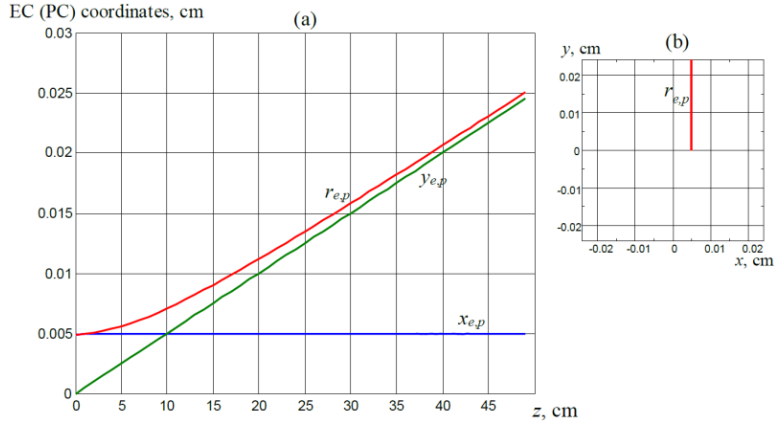

Fig. 2. (a) Variation of the EC (PC) coordinates (16) $x_{e,p}$ (blue), $y_{e,p}$ (green) and the EC (PC) off-axis shift $r_{e,p}$ (red) with growing distance from the beam waist $z$ for the monochromatic case $k_2 = k_1 = 10^5$ cm$^{-1}$. (b) View of the EC trajectory from the positive end of axis $z$.

Fig. 2 illustrates the EC and PC positions $x_{e,p}$, $y_{e,p}$; additionally, red curves show the behavior of $r_{e,p} = \sqrt{x_{e,p}^2 + y_{e,p}^2}$. Expectedly, in the monochromatic case, there is no difference between the EC and PC. With growing $z$, these evolve (propagate) along straight lines. Importantly, the spatial pattern of the monochromatic superposition is stationary (time-independent): the slow-varying terms discussed in Section 2 exactly vanish.

### 4.2. Specific features of the polychromatic MB

The general case of inequal wave numbers in Eqs. (11), (12) differs from the simple situation considered above in several important aspects [25,26]. When $k_1 \neq k_2$, even the time-averaged energy (probability) distributions are not stationary due to non-zero slow-varying terms (see

Section 2), and three special situations can be considered: (A) the pattern observed in a fixed cross section ($z =$ const) at varying time ($t =$ var); (B) the field pattern observed along the variable propagation distance ($z =$ var) at a fixed moment of time ($t =$ const); (C) the field pattern observed in the moving cross section "flying" together with the beam ($z = ct$). Of course, these three patterns are interrelated and reflect different sides of the combined (3+1)D picture of the MB evolution. Only the situation (C) can be directly applied as a model of the spatio-temporal wave packet propagation but the cases (A) and (B) expose some interesting aspects of the polychromatic paraxial beams and are therefore objects of special attention.

In a fixed cross section (situation (A)), the field pattern of the MB oscillates with the frequency $\Delta\omega$, and the intensity distribution looks as the asymmetric peak of Fig. 1 rotating around the axis $z$ (Fig. 3) with the angular velocity $\Delta\omega$. Accordingly, the EC and PC (asterisks in Fig. 3) are shifted from the axis $z$ and also rotate. However, in all numerical conditions considered, the probability distribution (8) visually looks very similar to the light energy distribution (5). To make their difference more sensible, in Fig. 3 the case of rather high $\Delta\omega = 0.4\omega_1$ is illustrated. Properly speaking, this case is marginal for our analysis because for such frequency difference one can hardly substantiate why the oscillation processes occurring with frequency $\Delta\omega$ are considered explicitly whereas the oscillations occurring with the sum frequency $\omega_1 + \omega_2 = 6\Delta\omega$ are time-averaged. For this reason, Fig. 3 (as well as Figs. 7 – 9 below) are mainly presented for the illustration purposes discovering some trends associated with high $\Delta\omega$ but are not directly applicable to realistic wave packets. Remarkably, Fig. 3 shows that spatial distributions of the energy and of the photon probability visually coincide even in this marginal case (however, their slight differences can be detected through the more detailed analysis, see Fig. 9, and the corresponding discrepancies in the EC and PC positions can be seen in Fig. 7).

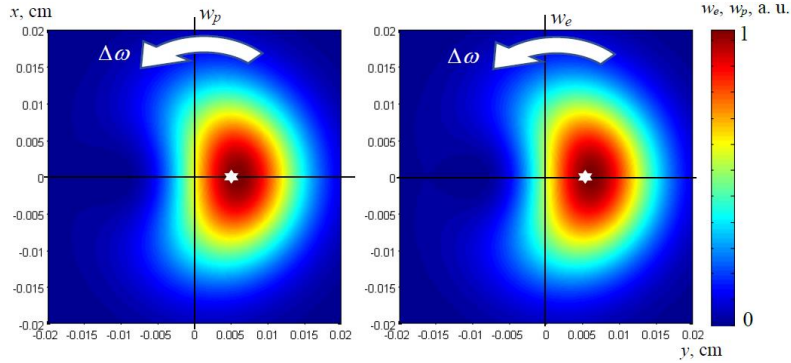

Fig. 3. Probability distribution (8) and energy distribution (5) in the section $z = 0$ at the time moment $t = 0$. With growing time, the pattern rotates counter-clockwise if $k_2 > k_1$ (in this figure, $k_1 = 10^5$ cm$^{-1}$, $k_2 = 1.4k_1$ were accepted), and clockwise if $k_2 < k_1$. Asterisks denote the EC and PC positions which are practically identical.

The rotating transverse intensity distribution enables to associate the considered MB with the family of rotating beams intensively studied in the past decades [24–27]. In particular, there is a specific behavior of the transverse momentum (18) and the corresponding energy flow distribution. In the initial plane ($z = 0$) it can be calculated using the explicit expressions (11), (12), and the result is time-dependent. For conditions (20) with real $A$ one easily obtains

$$P_x = \frac{A^2}{8\pi c k_2 b_0^3}\left[\frac{\Delta k}{k_1}\frac{xy}{b_0^2}\cos(\Delta\omega \cdot t) - \frac{\Delta k}{k_1}\frac{x^2}{b_0^2}\sin(\Delta\omega \cdot t) - \sin(\Delta\omega \cdot t) - \frac{y}{b_0}\right]\exp\left(-\frac{x^2+y^2}{b_0^2}\right),$$

$$P_y = \frac{A^2}{8\pi c k_2 b_0^3} \left[ \frac{\Delta k}{k_1} \frac{y^2}{b_0^2} \cos(\Delta\omega \cdot t) - \frac{\Delta k}{k_1} \frac{xy}{b_0^2} \sin(\Delta\omega \cdot t) + \cos(\Delta\omega \cdot t) + \frac{x}{b_0} \right] \exp\left(-\frac{x^2 + y^2}{b_0^2}\right). \quad (22)$$

Remarkably, the time-dependent terms of (22) give zero contributions to the total beam OAM (19), while the non-vanishing terms "generate" circular flow $\mathbf{e}_x P_x + \mathbf{e}_y P_y \propto -\mathbf{e}_x y + \mathbf{e}_y x = r\mathbf{e}_\phi$ ($r$, $\phi$ are the polar coordinates in the transverse cross section, $\mathbf{e}_\phi$ is the unit vector of the azimuthal direction), and the resulting OAM

$$L_z = \frac{A^2}{8\omega_2} = \frac{w_e}{\omega_2}$$

exactly coincides with the OAM of the beam (12) taken separately. This result is in contrast to the data of [25,26] where the OAM of rotating beams was proportional to the angular velocity of rotation and directed oppositely to it. This discrepancy can be understood regarding the fact that rotating beams considered in [25,26] were composed of oppositely charged vortex LG modes (with different frequencies) so that their resulting OAM vanished. In our MB, the OAM of the beam $u_2$ is not compensated and is fully imparted to the superposition $u_1 + u_2$ (11), (12). Accordingly, the OAM of the MB is not related with its rotational properties but originates from the "input" OAM of the vortex component $u_2$.

### 4.3. Longitudinal spatial pattern of the MB

At a fixed moment of time (situation (B)), one deals with the 3D optical-field spatial distribution, and can study its longitudinal evolution as it is usual in paraxial beams [28,29]. The spatial field distribution of the MB can be briefly characterized by the EC and PC. By using Eqs. (6), (9) and (16), one finds that these are positioned along helical lines spiraling around the propagation axis $z$ with the longitudinal period $z_T = 2\pi/|\Delta k|$. In this context, the situation of Fig. 2 with $\Delta k = 0.4 k_1$ is not favorable for visual demonstration of these variations because it determines the microscopic spiral period $z_T \approx 3.14$ μm. That is why Figs. 4 and 5 illustrate other circumstances with rather small $\Delta\omega$ where the spiral oscillations are macroscopic. Obviously, in these situations the difference between the EC and PC positions is vanishingly small and cannot be detected in the graphs.

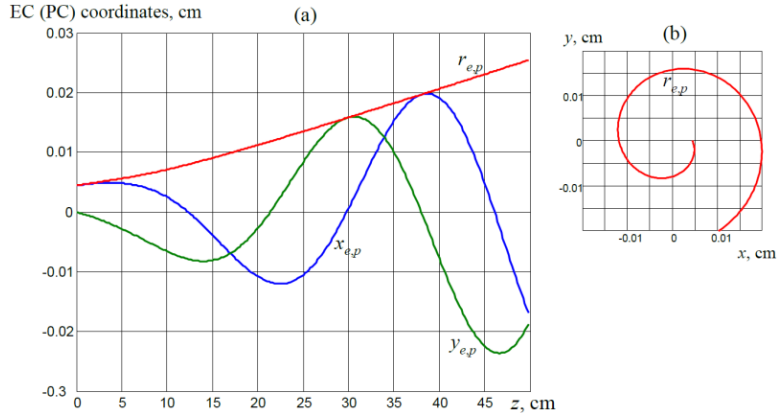

Fig. 4. (a) Longitudinal dependence of the EC (PC) coordinates $x_{e,p}$ (blue), $y_{e,p}$ (green), and $r_{e,p}$ (red) for $k_2 = k_1 + 0.2$ cm$^{-1}$ and $t = 0$ (period of oscillations 31.4 cm; the range of $z$ (0, $5z_R$) includes ~1.6 periods). (b) View of the EC trajectory from the positive end of axis $z$.

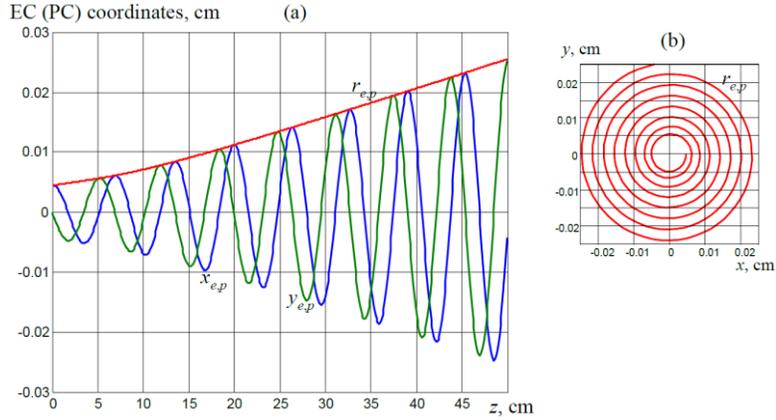

Fig. 5. (a) Longitudinal dependence of the EC (PC) coordinates $x_{e,p}$ (blue), $y_{e,p}$ (green), and $r_{e,p}$ (red) for $k_2 = k_1 + 1$ cm$^{-1}$ and $t = 0$ (period of oscillations 6.28 cm, the same range of $z$ as in Fig. 3 includes ~8 periods). (b) View of the EC trajectory from the positive end of axis $z$.

### 4.4. Flying section

In the "flying section" situation (C), $s = z - ct = 0$, and this condition is the most relevant for the "traveling wave packet" model. At $s = 0$, the terms of expressions (5), (8), (16), (18) describing the "slow" oscillations with the frequencies $\Delta\omega$ and $\Delta k$ disappear: the field evolves monotonically, and the EC and PC trajectories in Fig. 6 look, at first glance, rectilinear, as in the stationary case of Fig. 2. However, with growing propagation distance, the difference in spatial parameters (13) of the superposing MB components (11) and (12) is "accumulated": e.g., while initially (at $z = t = 0$) the beam waist parameters $b_0$ were identical for both components (11) and (12), their increments after the propagation distance $z$ differ due to the difference of $k_1$ and $k_2$ in (13). Accordingly, the EC (PC) trajectories are no longer rectilinear as in Fig. 2; the curvilinear deviations occur in Fig. 6. For small $\Delta\omega$, these deviations are only seen at rather large distances $z$ but with growing $\Delta\omega$, they become noticeable at smaller $z$ and are expressed more evidently (cf. thick and thin curves in Fig. 6).

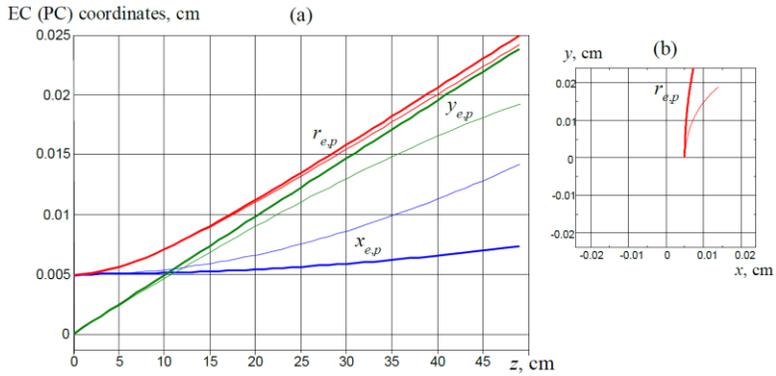

Fig. 6. (a) Evolution of the EC and PC coordinates $x_{e,p}$ (blue), $y_{e,p}$ (green), and $r_{e,p}$ (red) in the cross section $z = ct$ moving from the common waist of beams (11), (12) for $k_1 = 10^5$ cm$^{-1}$, $k_2 = 1.02k_1$ (thick lines) and $k_2 = 1.1k_1$ (thin lines). (b) View of the EC and PC trajectories from the positive end of axis $z$.

Nevertheless, the EC and PC positions are still indistinguishable in the curves of Fig. 6. To make their difference visible, we again resort to the "marginal" conditions $k_2 = 1.4k_1$ (cf. Fig. 3 and the related comments). The results are presented in Fig. 7, and they testify that even for such a huge non-monochromaticity, the EC and PC positions differ by nearly 1%, which is negligible in many cases. This fact is confirmed by a direct evaluation via Eq. (17) which gives $x_e/x_p = y_e/y_p = 1.014$ for $k_2 = 1.4k_1$ and $A_1 = A_2$ (see Eq. (20)). This example supports the general conclusion, at first glance counter-intuitive, that the EC and PC of paraxial wave packets coincide with a high accuracy almost in all practically meaningful situations [30].

Simultaneously, Fig. 7 spectacularly shows another effect of the large frequency difference of the MB components (11) and (12): deviations of the curves $x_{e,p}(z)$, $y_{e,p}(z)$ from the rectilinear prototypes characteristic for the monochromatic superposition (see Fig. 2) become very impressive.

In general, the rectilinear EC and PC evolution (similarity to Fig. 2) follows from Eq. (5) whose form for $s = 0$ is the same as for $\Delta k = 0$. Accordingly, the intensity distribution in the flying section of the bi-chromatic MB (Fig. 8) evolves similarly to the transverse beam pattern of the monochromatic superposition (Fig. 1), at least until the distance from the common waist plane is not very high (first row of Fig. 8). However, with growing distance $z$, the difference in spatial parameters (13) of the MB components leads to the beam profile distortion well seen in the marginal case $\Delta k = 0.4k_1$ (2nd and 3rd rows of Fig. 8); it is this distortion that is responsible for the deviations from linear $x_{e,p}(z)$ and $y_{e,p}(z)$ evolutions in Figs. 6 and 7.

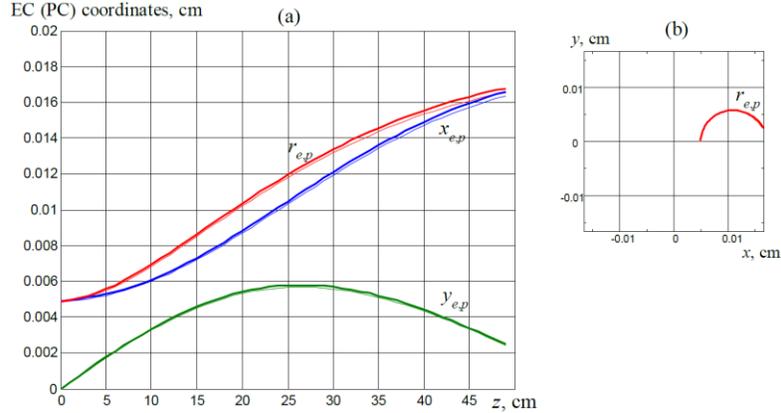

Fig. 7. (a) Evolution of the EC (thick lines) and PC (thin lines) coordinates $x_{e,p}$ (blue), $y_{e,p}$ (green), and $r_{e,p}$ (red) in the cross section $z = ct$ moving from the common waist of beams (11), (12) for $k_1 = 10^5$ cm$^{-1}$, $k_2 = 1.4k_1$. (b) View of the EC and PC trajectories from the positive end of axis $z$.

The especially impressive illustration of this distortion can be seen at high propagation distances $z > 5z_R$ (bottom row of Fig. 8). There, one can see that, while the bright spot approaches the positive $y$-semiaxis, its bottom-right periphery gradually elongates and forms the spiral-like "tail". Herewith, the sensible portion of the light energy moves clockwise to the $x$-axis and further; this is the reason why $y_{e,p}(z)$ curves deviate from the rectilinear form (cf. the thick green curves in Figs. 2, 6 and the thin green line of Fig. 6) and, ultimately, turn back to the horizontal axis (thick and thin green curves in Fig. 7, red curve in Fig. 7b). It is not shown in figures but with further growth of $z > 50$ cm, the curve $y_{e,p}(z)$ may reach the horizontal axis and even cross it. This effect is quite similar to the deformation of a rotating beam profile reported previously (see Fig. 5 in [25]). The only difference is that in [25], the superposition with rather low $\Delta k = 10^{-5}$ cm$^{-1}$ was discussed, and, therefore, the beam profile distortion becomes only noticeable at very high propagation distances. As in [25], effects of the "flying section" profile deformation become observable at propagation distances higher than $z \approx (k_1 / \Delta k) z_{R1}$, which equals to 25 cm upon conditions of Figs. 7, 8.

Our last task is to discuss explicitly the difference between the energy and probability distributions. With the help of the MB (11), (12), this can be performed numerically if the conditions of high frequency difference are accepted. Fig. 9 illustrates the relative discrepancy between the distributions (5) and (8), defined as

$$\Delta_{pe}(x,y) = \frac{w_p(x,y) - w_e(x,y)}{\max(w_p)}, \qquad (23)$$

calculated for the MB (11), (12) upon the conditions (20) for different $\Delta k$ values. One can see that both distributions are very similar with high accuracy, and even at a large non-monochromaticity ($\Delta k = 0.1 k_1$, left image of Fig. 9), their relative disagreement does not exceed 2%. Besides, the maximum relative discrepancy (23) takes place near the axis ($x = y = 0$) and is spatially separated from the maxima of both $w_p$ and $w_e$ (cf. Fig. 3), so the region with relatively large $\Delta_{pe}$ makes a relatively small contribution to the integral characteristics EC and PC (6), (9), (16). Even the large local difference between $w_p(x,y)$ and $w_e(x,y)$ (observable, e.g., at $k_2 = 1.4k_1$ in Fig. 9) rather slightly affects the difference between the EC and PC positions, which, according to Eq. (16), is mainly determined by the integral quantities (7) and (10). This observation provides additional qualitative argumentation why the EC and PC positions coincide with a high accuracy in most cases [30].

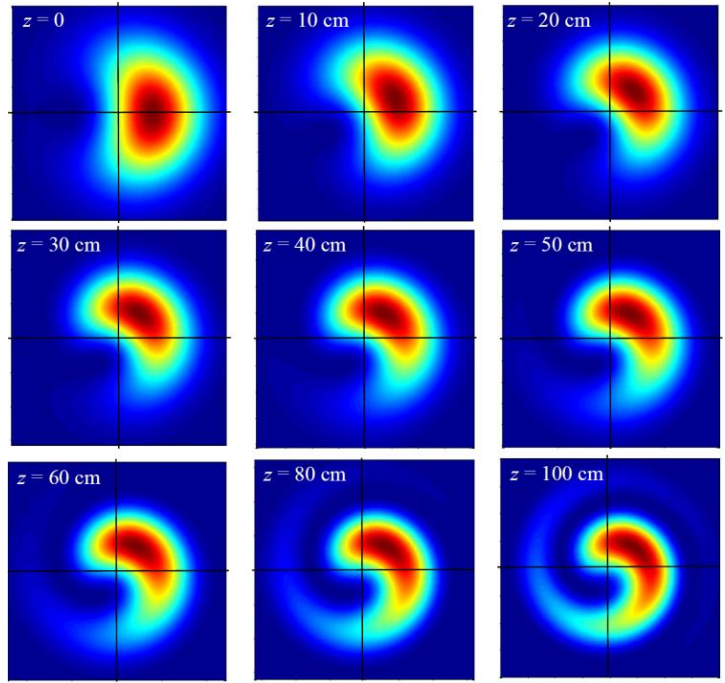

Fig. 8. Transverse intensity distributions of the MB in the "flying" cross section $z = ct$ at various distances from the common waist of beams (11), (12) for $k_1 = 10^5$ cm$^{-1}$, $k_2 = 1.4 k_1$; sizes of the images are $2b_1(z) \times 2b_1(z)$, $b_1(z)$ is determined by Eq. (13).

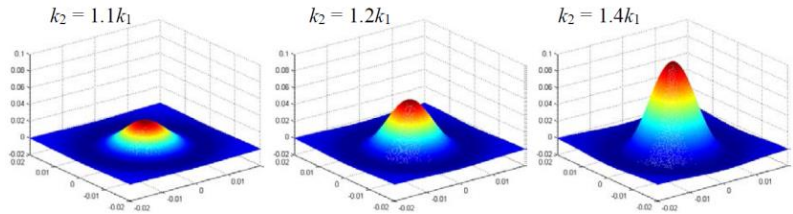

Fig. 9. Normalized discrepancy (23) between the transverse energy (5) and probability (8) distributions in the initial cross section $z = 0$ at the time moment $t = 0$, for various values of the wavenumber difference $\Delta k$.

An important remark can be made on the relevancy of the beam profile distortion described by Fig. 8 for the realistic spatio-temporal pulses modeled by the MB. Obviously, the influence of the non-monochromatic contributions on the MB structure is enhanced by the fact that these are concentrated in the "boundaries" of the spectral range separated by the interval $\Delta k$. In real ultra-short pulses, the spectral "boundaries" carry relatively small part of the light energy, and typically, the monochromatic components with much smaller frequency differences interfere; additionally, the limited length of a pulse prevents from accumulation of essential discrepancies in the spatial characteristics of its monochromatic components. As a result, the transverse beam pattern distortions would be much less expressive, or practically negligible, in realistic polychromatic superpositions.

## 5. Conclusion

In this paper, we consider the superpositions of two monochromatic beams with different frequencies and inspect how such bi-chromatic model beams (MB) manifest general features of realistic spatio-temporal wave packets. The results can be useful for heuristic discussion and understanding of mechanisms underlying the formation and evolution of non-monochromatic light structures.

Of course, the MB provides a rather coarse model of real situations. The MB is infinite in time and in the longitudinal dimension $z$ whereas the real light pulse has limited length and duration; within these dimensions, the MB is characterized by the spiral quasi-periodic structure, which is not typical for the modeled spatio-temporal fields. The numerical predictions of the described model are sometimes hardly reliable, and this is its drawback, which, however, is directly related with its main merit of simplicity. The periods of the MB spatial and temporal behavior give a very approximate impression on the light-pulse length and duration; nevertheless, the MB "flying" section, propagating with the light velocity $c$ (Section 4.4), supplies a relatively adequate representation of a realistic spatio-temporal pulse propagating along the $z$-direction.

Remarkably, the paraxial bi-chromatic model permits to elucidate some principal problems associated with general polychromatic wave packets. In particular, it enables an explicit, pictorial and self-consistent analysis of the photon-probability distribution, whose general description in spatio-temporal light fields meets difficulties associated with its ill-definiteness in the coordinate space [21]. Based on the simple examples introduced in this paper, the probability density is considered in a simple and direct way, disclosing its important physical features, similarities and discrepancies with the energy density, numerical proximity of the energy center (EC) and probability center (PC), etc.

It should be emphasized that in this paper, adhering the paraxial-optics framework, only the transverse $(x, y)$-sections of the energy and probability distributions are considered, and for these sections the EC and PC are defined [Eqs. (6), (9), (16)]. Due to these "local" definitions, the EC and PC show the rich and intricate spatio-temporal evolution illustrated by Figs. 4 – 7. In real 3D wave packets, the EC and DC determined for the whole field, including the longitudinal coordinate, move rectilinearly, obeying the free-space propagation law [8,23,30]. This property is shared by the "flying section" of the MB, which additionally supports its particular modeling appropriateness mentioned in the above paragraphs.

Finally, we briefly discuss possible developments of the proposed model. The MB in the form of superposition of the LG beams (11) and (12) was chosen deliberately; in fact, any other two-frequency superposition could be employed, and this freedom allows one to improve the model and adapt it better to any realistic spatio-temporal field. This circumstance excites additional interest to the similar non-monochromatic superpositions considered earlier and classified as "rotating light beams" [25–27]. In this context, the MB introduced in this paper can be considered as an element of a new subclass of rotating beams, whose investigation constitutes an independent interest. Its main distinction from the previous examples of [25,26] is that the OAMs of the components (11), (12) are not compensated, and the total OAM of the MB is not related with the visible rotation of the beam profile around the propagation axis. At the same time, the possibility of controllable rotation of the intensity maximum or minimum within a chosen cross section prompts interesting applications of the rotating beams of this type for the optical trapping and manipulation techniques.

**Funding.** Ministry of Education and Science of Ukraine, project No 610/22, state reg. # 0122U001830.

**Acknowledgments.** The author is grateful to K. Bliokh (Theoretical Quantum Physics Laboratory, RIKEN Cluster for Pioneering Research, Wako-shi, Saitama 351-0198, Japan) for fruitful discussion.

**Disclosures.** The author declares no conflicts of interest.

**Data availability.** Data underlying the results presented in this paper are not publicly available at this time but may be obtained from the authors upon reasonable request.